\documentclass[aps,preprint,showpacs,floatfix]{revtex4-1}
\usepackage{bm}
\usepackage{amsmath}
\usepackage{epsfig}
\usepackage{graphicx}
\usepackage{subfigure}

\def\pr{\prime}
\def\be{\begin{equation}}
\def\lan{\left\langle}
\def\ran{\right\rangle}
\def\ee{\end{equation}}
\def\barr{\begin{array}}
\def\earr{\end{array}}

\def\nn8{\\}
\def\l{\left}
\def\r{\right}
\def\dis{\displaystyle}
\def\ed{\end{document}}

\def\dg{\dagger}

\def\cac{{\cal C}}

\def\capp{{\cal P}}

\def\cs{{\bf s}}

\def\we{{\widehat {E}}}

\def\wh{{\widehat {H}}}
\def\whh{{\widehat {h}}}
\def\wv{{\widehat {V}}}

\begin{document}

\title{Random matrix ensemble with random two-body interactions in presence of a
mean-field for spin one boson systems}

\author{H. N. Deota$^{1}$, N. D. Chavda$^{1,}\,$\footnote{Corresponding author,
Phone: +91 265 2434188 EXT. 211, Fax: +91 265 2423898 \\ {\it E-mail address:} ndchavda-apphy@msubaroda.ac.in (N.D. Chavda)
}, V. K. B. Kota$^{2}$, V.Potbhare$^{1}$ and Manan Vyas$^{2,3}$}

\affiliation{$^1$Applied Physics Department, Faculty of Technology and
Engineering, M.S. University of Baroda, Vadodara 390 001, India \\ $^2$Physical
Research  Laboratory, Ahmedabad 380 009, India\\ $^3$ Department of Physics and
Astronomy, Washington State University, Pullman WA 99164-2814  }

\begin{abstract}

For $m$ number of bosons, carrying spin ($S$=$1$) degree of freedom, in $\Omega$
number of single particle orbitals, each triply degenerate, we introduce and
analyze embedded Gaussian orthogonal ensemble of random matrices generated by
random two-body interactions that are spin (S) scalar [BEGOE(2)-$S1$]. The
embedding algebra is $U(3) \supset G \supset G1 \otimes SO(3)$ with $SO(3)$
generating spin $S$. A method for constructing the ensembles in fixed-($m$, $S$)
space has been developed. Numerical calculations show that the form of the
fixed-($m$, $S$) density of states is close to Gaussian and level fluctuations
follow GOE. Propagation formulas for the fixed-($m$, $S$) space energy centroids
and spectral variances are derived for a general one plus two-body Hamiltonian
preserving spin. In addition to these, we also introduce two different pairing
symmetry algebras in the space defined by BEGOE(2)-$S1$ and the structure of
ground states is studied for each paring symmetry.

\end{abstract}

\pacs{05.30.Jp, 05.30.-d, 03.65.Aa, 03.75.Mn, 21.60.Fw}

\maketitle

\section{Introduction}
\label{sec:s1}
Embedded Gaussian orthogonal ensembles of one- plus two-body interactions for
finite isolated interacting spin-less many boson systems  [denoted by
BEGOE(1+2)] were introduced and studied in detail in the last decade in
\cite{Pa-00,Ag-01,Ch-03,Ch-04,Ch-arxiv}. Going beyond spin-less boson systems,
very recently embedded Gaussian orthogonal ensemble of random matrices for two
spices boson systems with $F$-spin degree of  freedom for Hamiltonians that
conserve the total $F$-spin of the $m$-boson systems [called BEGOE(1+2)-$F$] is
introduced and  its spectral properties are analyzed in detail in
\cite{vyas-12}; the $F$-spin for the bosons is similar to the $F$-spin in the
proton-neutron interacting boson model  (pnIBM) of atomic nuclei \cite{Ca-05}.
Another interesting extension of BEGOE  is to a system of bosons carrying spin
one ($S=1$) degree of freedom. With random two-body interactions preserving many
boson spin $S$ then generates the ensemble called hereafter BEGOE(2)-$S1$. In
the presence of a mean-field the corresponding ensemble is BEGOE(1+2)-$S1$. The
purpose of the present paper is to introduce this ensemble and report results of
the first analysis, both numerical and analytical, of this ensemble.
BEGOE(1+2)-$S1$ ensemble will be useful for spinor BEC discussed in
\cite{Pe-10,PRA} and in the analysis of IBM-3 model of atomic nuclei (here spin
$S$ is isospin $T$ of the bosons in IBM-3) \cite{Ga-99,Ko-98}. Moreover, there
is a considerable interest in analyzing a variety of embedded ensembles as they
can be used as generic models for many-body chaos \cite{Go-11,MW-10} and hence
 useful to analyze BEGOE(1+2)-$S1$ ensemble. Now we will give a preview.

In Section \ref{sec:s2}, introduced is the embedded ensemble BEGOE(1+2)-$S1$ [also
BEGOE(2)-$S1$] for a system of $m$ bosons in $\Omega$ number of sp orbitals that
are triply degenerate with total $S$-spin being a good symmetry. A method for
the numerical construction of this ensemble in fixed-($m$, $S$) space is
described. In Section \ref{sec:s3}, embedding algebra, $U(\Omega) \otimes [SU(3) \supset
SO(3)]$ for BEGOE(1+2)-$S1$  is described. Section \ref{sec:s4} contains some numerical results
for the ensemble  averaged eigenvalue density, nearest neighbor spacing
distribution (NNSD), width of the fluctuations in energy centroids and spectral
variances. In addition, propagation formula for fixed-($m$,$S$)
energy centroids for general one- plus two-body Hamiltonians that preserve $S$
and a method to propagate the spectral variances are given. In Section \ref{sec:s5}, two
types of pairing in BEGOE(1+2)-$S1$ space are introduced and some numerical
results for ground state structure vis-a-vis the two different  pairing
interactions are presented. Finally, Section \ref{sec:s6} gives conclusions and future
outlook.

\section{Definition and construction}
\label{sec:s2}
Let us consider a system of $m$ ($m>2$) bosons with spin 1 ($S=1$) degree of
freedom and occupying $\Omega$ number of sp levels. For convenience, in the
remaining part of this section, we use the notation $\cs$ for the spin quantum
number of a single boson, $s$ for the spin carried by a two boson  system and
for $m>2$ boson systems $S$ for the spin. Therefore, $\cs=1$; $s= 0$, $1$ and
$2$; $S=m$, $m-1$, $\ldots$, $0$. Similarly, the $\hat{S}_z$ ('hat' denoting
operator) eigenvalue is denoted by $m_\cs$, $m_s$ and $M_S$ respectively. Now
on, the space generated by the sp levels $i=1,2$, $\ldots$, $\Omega$ is referred
as orbital space. Then the sp states of a boson are denoted by  $\l.\l| i;
\cs=1,m_\cs \r.\ran$ with $i=1,2,\ldots,\Omega$ and $m_\cs=+1$, $0$ and $-1$.
With $\Omega$ number of orbital degrees of freedom and three spin ($m_\cs$)
degrees of freedom, total number of sp states is $N=3\Omega$. Going further, two
boson (normalized) states that are symmetric in the total orbital $\times$ spin
space are denoted by $\l.\l|(ij); s,m_s\r.\ran$ with $s= 1 \times 1= 0$, $1$ and
$2$; however, for $i=j$ only $s=0,2$ are allowed.

For one plus two-body Hamiltonians preserving $m$-particle spin $S$, the
one-body Hamiltonian $h(1)$ is defined by the sp energies $\epsilon_i$;
$i=1,2,\ldots,\Omega$,   with average spacing $\Delta$,
\be
\whh(1)=\dis\sum_{i=1}^{\Omega}\epsilon_i \hat{n}_i
\label{eq.bege-p1}
\ee
where $\hat{n}_i=\sum_{m_\cs} \hat{n}_{i:m_\cs}$ counts number of bosons in the
$i$-th orbit. Similarly the two-body Hamiltonian $V(2)$ is defined by the
two-body matrix elements $V^s_{ijkl}(2)=\lan (kl)s,m_s \mid \wv(2) \mid (ij)
\r.$ $\l. s,m_s\ran$  with the  two-particle spin $s$ taking values $0$, $1$ and
$2$. These matrix elements are independent of  the $m_s$ quantum number.  The
$V(2)$ matrix in two-particle space will be a direct sum three matrices
generated by the three $\wv^s(2)$ operators respectively. Now the
BEGOE(1+2)-$S1$ Hamiltonian is
\be
\l\{\wh(1+2)\r\}= \whh(1) + \lambda_0
\l\{\wv^{s=0}(2)\r\} + \lambda_1 \l\{\wv^{s=1}(2)\r\} +
\lambda_2 \l\{\wv^{s=2}(2)\r\}
\label{eq.bege-p2}
\ee
with three parameters $(\lambda_0,\lambda_1,\lambda_2)$.  Now, BEGOE(2)-$S1$
ensemble for a given $(m,S)$ system is generated  by defining the three parts of
$\wv(2)$ in two-particle space to be independent GOE(1)s [i.e.,  matrix
elements are independent Gaussian variables with zero center and variance unity
for off-diagonal matrix elements and 2 for diagonal matrix elements] and then
propagating each member of the $\l\{\wh(1+2)\r\}$ to the $m$-particle space
with a given spin $S$ by using the geometry (direct product structure) of  the
$m$-particle space. A method for carrying out the propagation is discussed
ahead. With $\whh(1)$ given by Eq. (\ref{eq.bege-p1}), the sp levels will be
triply degenerate with average spacing $\Delta$. Without loss of generality we
put $\Delta=1$ so that the $\lambda$s in Eq. (\ref{eq.bege-p2})  will be in
units of $\Delta$.

For generating a many-particle basis, firstly, the sp states are arranged such
that the first $\Omega$ number of sp states have $m_\cs = 1$,   next $\Omega$
number of sp states have $m_\cs = 0$ and the remaining $\Omega$ sp states have
$m_\cs = -1$. Now, the many-particle states for $m$ bosons can be obtained by
distributing $m_1$ bosons in the $m_\cs = 1$ sp states,  $m_2$ bosons in the
$m_\cs = 0$ sp states and similarly, $m_3$ bosons in the $m_\cs = -1$ sp
states with $m=m_1+m_2+m_3$.  Thus, $M_S = (m_1 -m_3)$. Let us denote each
distribution of $m_1$ bosons in $m_\cs = 1$ sp states by $\bf{m}_1$, $m_2$
bosons in $m_\cs = 0$ sp states by $\bf{m}_2$ and similarly,  $\bf{m}_3$ for
$m_3$ bosons in  $m_\cs = -1$ sp states. Configurations defined by $(\bf{m}_1,
\bf{m}_2, \bf{m}_3)$  will form a basis for constructing $H$ matrix in $m$-boson
space. Action of the  Hamiltonian operator defined by Eq. (\ref{eq.bege-p2})
on  $(\bf{m}_1, \bf{m}_2, \bf{m}_3)$ basis states with fixed-($m,M_S=0$)
generates the ensemble in ($m,M_S$) space. It is important to note that the
construction of the $m$-particle $H$ matrix in fixed-($m,M_S=0$) space reduces
to the  problem of BEGOE(1+2) for spinless boson systems and hence Eq.
(4) of \cite{Pa-00} will apply. For this, we need to convert the $H$ operator
into $M_S$ representation. Two boson states in $M_S$ representation can be
written as $\l| i, m_\cs; j, m^{\pr}_\cs \ran$; $m_s=m_\cs + m^{\pr}_\cs$.
The the two particle matrix elements are $V^{\pr}_{i,m^{f1}_\cs ; j,m^{f2}_\cs
; k,m^{i1}_\cs; \ell, m^{i2}_\cs}(2) = \lan i,m^{f1}_\cs ; j,m^{f2}_\cs \mid
\wv(2) \mid k,m^{i1}_\cs; \ell, m^{i2}_\cs \ran$. It is easy to apply angular
momentum algebra and derive formulas for these in terms of  $V^s_{ijkl}(2)$. The
final formulas are,
\be
\barr{rcl}
V^{\pr}_{i,1;j,1;k,1;\ell,1}(2) & = & V^{s=2}_{ijkl}(2)\;, \\
\\
V^{\pr}_{i,1;j,0;k,1;\ell,0}(2) & = & \dis\frac{\dis\sqrt{(1+\delta_{ij})
(1+\delta_{k \ell})}}{2}\,\l[V^{s=1}_{ijkl}(2) + V^{s=2}_{ijkl}(2)\r]\;,\\
\\
V^{\pr}_{i,1;j,-1;k,1;\ell,-1}(2) & = & \dis\frac{\dis\sqrt{(1+\delta_{ij})
(1+\delta_{k \ell})}}{6}\,\l[2\,V^{s=0}_{ijkl}(2) + 3\,V^{s=1}_{ijkl}(2) +
V^{s=2}_{ijkl}(2) \r]\;,\\
\\
V^{\pr}_{i,0;j,0;k,0;\ell,0}(2) & = & \l[\dis\frac{1}{3}\,V^{s=0}_{ijkl}(2) +
\dis\frac{2}{3}\,V^{s=2}_{ijkl}(2)\r] \;, \\
\\
V^{\pr}_{i,1;j,-1;k,0;\ell,0}(2) & = & \dis\frac{\dis\sqrt{(1+\delta_{ij})}}{3}
\,\l[V^{s=2}_{ijkl}(2) - V^{s=0}_{ijkl}(2)\r]\;.
\earr \label{eq.bege-p3}
\ee
All other $V^{\pr}$ matrix elements follow by symmetries. The fact that the sp
energies $\epsilon$ are independent of $m_\cs$, Eq. (\ref{eq.bege-p3}) above and
Eq. (4) of \cite{Pa-00} will allow one to construct the $H$-matrix in $(\bf{m}_1,
\bf{m}_2, \bf{m}_3)$ basis for a given value of $m$ and $M_S=0$.  Then, ${\hat
S}^2$ operator is used for projecting states with good $S$, i.e. to covert the
$H$-matrix into direct sum of matrices with block matrices for each allowed $S$
value. Matrix elements of ${\hat S}^2$ in $s=0$, $1$ and $2$ space are $-4$,
$-2$ and $2$ respectively. This procedure has been implemented and computer
programmes are developed. Some numerical results obtained using these programmes
will be discussed in Section \ref{sec:s3}. Let us add that the BEGOE(1+2)-$S1$ ensemble
is defined by five parameters $(\Omega, m, \lambda_0, \lambda_1, \lambda_2)$
with $\lambda_s$ in units of $\Delta$.

\section{$U(\Omega) \otimes [SU(3) \supset SO(3)]$ embedding algebra}
\label{sec:s3}

Embedding algebra for BEGOE(1+2)-$S1$ is not unique and following the earlier
results for the IBM-3 model of atomic nuclei \cite{Ga-99,Ko-98}, it is possible
to identify two algebras. They are: (i) $U(3\Omega) \supset U(\Omega)  \otimes
[U(3) \supset SO(3)]$; (ii) $U(3\Omega) \supset SO(3\Omega)  \supset SO(\Omega)
\otimes SO(3)$. Here we will consider (i) and later in Section \ref{sec:s5} we will
consider briefly (ii).

Firstly, the spectrum generating algebra $U(3\Omega)$ is generated by the
$(3\Omega)^2$ number of operators $u^k_q(i,j)$ where
\be
u^k_q(i,j) = \l(b^{\dagger}_{i;\cs=1} \tilde{b}_{j;\cs=1}\r)^k_q\;;
k=0,1,2\;\;\mbox{and}\;\;i,j=1,2,\ldots,\Omega\;.
\label{eq.bege-p4}
\ee
Note that $u^k$ are given in angular momentum coupled representation with $k=
\cs \times \cs=0,1,2$. Also, $b^\dg$ are one boson creation operators, $b$ are
one boson annihilation operators and $\tilde{b}_{i;1,m_\cs} =
(-1)^{1+m_\cs}\,b_{i;1,-m_\cs}$. The quadratic Casimir invariant of
$U(3\Omega)$ is
\be
\hat{C}_2(U(3\Omega))=\dis\sum_{i,j,k} u^k(i,j) \cdot u^k(j,i)\;.
\label{eq.bege-p5}
\ee
Note that $T^k \cdot U^k = (-1)^k \sqrt{(2k+1)}\,(T^k U^k)^0$. In terms
of the number operator $\hat{n}$,
\be
\hat{n} = \dis\sum_{i,m_\cs} b^\dagger_{i;1,m_\cs} b_{i;1,m_\cs}\;,
\label{eq.bege-p6}
\ee
we have
\be
\hat{C}_2(U(3\Omega))= \hat{n} (\hat{n}+ 3\Omega -1)\;.
\label{eq.bege-p7}
\ee
All $m$-boson states transform as the symmetric irrep $\{m\}$ w.r.t.
$U(3\Omega)$ algebra and
\be
\lan \hat{C}_2(U(3\Omega)) \ran^{\{m\}} = m(m+3\Omega-1) \;.
\label{eq.bege-p8}
\ee
Using the results given in \cite{Ko-00} it is easy to write the generators of
the algebras $U(\Omega)$ and $SU(3)$ in $U(3\Omega) \supset U(\Omega) \otimes
SU(3)$. The $U(\Omega)$ generators are $g(i,j)$ where,
\be
g(i,j)= \dis\sqrt{3}\,\l(b^{\dagger}_{i;\cs=1} \tilde{b}_{j;\cs=1}\r)^0\;;
i,j=1,2,\ldots,\Omega
\label{eq.bege-p9}
\ee
and they are $\Omega^2$ in number. Similarly, $SU(3)$ algebra is generated by
the eight operators $h^{k=1,2}_q$ where,
\be
h^k_q = \dis\sum_i\,\l(b^{\dagger}_{i;\cs=1} \tilde{b}_{i;\cs=1}\r)^k_q\;;
k=1,2\;.
\label{eq.bege-p10}
\ee
It is useful to mention that ($h^0$, $h^1_q$, $h^2_{q^\pr}$) generate $U(3)$
algebra and $U(3) \supset SU(3)$. The quadratic Casimir invariants of
$U(\Omega)$ and $SU(3)$ algebras are,
\be
\barr{rcl}
\hat{C}_2(U(\Omega)) & = & \dis\sum_{i,j}\;g(i,j) \cdot g(j,i)\;,\\
\hat{C}_2(SU(3)) & = & \dis\frac{3}{2} \dis\sum_{k=1,2} k^k \cdot h^k\;.
\earr \label{eq.bege-p11}
\ee
The irreps of $U(\Omega)$ can be represented by Young tableaux $\{f\}=\{f_1,
f_2,\ldots,f_\Omega\}$, $\sum_i f_i = m$. However, as we are dealing with boson
systems (i.e. the only allowed $U(3\Omega)$ irrep being $\{m\}$), the irreps of
$U(\Omega)$ and $U(3)$ should be represented by the same $\{f\}$. Therefore,
$\{f\}$ will be maximum of three rows. The $U(\Omega)$ and $SU(3)$ equivalence
gives a relationship between their quadratic Casimir invariants,
\be
\barr{rcl}
\hat{C}_2(U(\Omega)) & = & \hat{C}_2(U(3)) + (\Omega-3)\,\hat{n} \;,\\
\hat{C}_2(U(3)) & = & \dis\sum_{k=0,1,2} h^k \cdot h^k = \dis\frac{2}{3}
\hat{C}_2(SU(3)) + \dis\frac{\hat{n}^2}{3}\;.
\earr \label{eq.bege-p11-a}
\ee
These relations are easy to prove using Eqs. (\ref{eq.bege-p9})-
(\ref{eq.bege-p11}). Given the $U(\Omega)$ irrep
$\{f_1\,f_2\,f_3\}$, the corresponding $SU(3)$ irrep in Elliott's notation
\cite{Ell-58} is given by $(\lambda \mu)$ where $\lambda=f_1-f_2$ and
$\mu=f_2-f_3$. Thus,
\be
\barr{l}
\{m\}_{U(3\Omega)} \rightarrow \l[\l\{f_1\,f_2\,f_3\r\}_{U(\Omega)}\r]\;
\l[(\lambda\,\mu)_{SU(3)}\r]\;;\\
f_1+f_2+f_3=m,\;\;\;f_1 \geq f_2 \geq f_3 \geq 0\;,\\
\lambda=f_1-f_2,\;\;\mu=f_2-f_3\;.
\earr \label{eq.bege-p12}
\ee
Using Eq. (\ref{eq.bege-p12}) it is easy to write, for a given $m$, all the
allowed $SU(3)$ and equivalently $U(\Omega)$ irreps. Eigenvalues of
$\hat{C}_2(SU(3))$ are given by
\be
\lan \hat{C}_2(SU(3)) \ran^{(\lambda\,\mu)} = C_2(\lambda\,\mu) =
\l[\lambda^2 + \mu^2 + \lambda \mu + 3(\lambda + \mu)\r]\;.
\label{eq.bege-p13}
\ee
Let us add that the $SU(3)$ algebra also has a cubic invariant $C_2(SU(3))$ and
its matrix elements are \cite{DR-NPA},
\be
\lan \hat{C}_3(SU(3)) \ran^{(\lambda\,\mu)} = C_3(\lambda\,\mu) =
\dis\frac{2}{9}\,(\lambda - \mu)(2\lambda +\mu +3)(\lambda +2\mu +3)\;.
\label{eq.bege-p13-1}
\ee
The $SO(3)$ subalgebra of $SU(3)$ generates spin $S$. The spin generators are
\be
S^1_q = \dis\sqrt{2}\; h^1_q\;,\;\;\; \hat{S}^2=C_2(SO(3))=S^1 \cdot S^1,
\;\;\;\lan \hat{S}^2 \ran^S = S(S+1)\;.
\label{eq.bege-p14}
\ee
Given a $(\lambda\,\mu)$, the allowed $S$ values follow from Elliott's rules
\cite{Ell-58} and this introduces a '$K$' quantum number,
\be
\barr{rcl}
K & = & min(\lambda\,,\,\mu),\; min(\lambda\,,\,\mu)-2,\;\ldots,\;0\;\;
\mbox{or}\;\;1\;,\\
S & = &  max(\lambda\,,\,\mu),\; max(\lambda\,,\,\mu)-2,\;\ldots,\;0\;\;
\mbox{or}\;\;1\;\;\mbox{for}\;\;\;K=0,\\
& =& K, K+1, K+2,\ldots,K+max(\lambda\,,\,\mu)\;\;\mbox{for}\;\;\;K \neq 0\;.
\earr \label{eq.bege-p15}
\ee
Eq. (\ref{eq.bege-p15}) gives $d_{(\lambda\,\mu)}(S)$, the number of times a
given $S$ appears in a $(\lambda\,\mu)$ irrep. Similarly the number of
substates  that belong to a $U(\Omega)$ irrep $\{f_1\, f_2\, f_3\}$ are given by
$d_{\Omega}(f_1\, f_2\, f_3)$ where \cite{Ko-06},
\be
d_{\Omega}(f_1\, f_2\, f_3) = \l| \barr{ccc} d_{\Omega}(f_1) &
d_{\Omega}(f_1 +1) & d_{\Omega}(f_1 +2) \\  d_{\Omega}(f_2 -1) &
d_{\Omega}(f_2) & d_{\Omega}(f_2 +1) \\
d_{\Omega}(f_3 -2) & d_{\Omega}(f_3 -1) & d_{\Omega}(f_3) \earr\r|\;.
\label{eq.bege-p16}
\ee
Here, $d_{\Omega}(\{g\})={\Omega+g-1 \choose m}$ and $d_{\Omega}(\{g\})=0$  for
$g < 0$. Note that the determinant in Eq. (\ref{eq.bege-p16}) involves only
symmetric $U(\Omega)$ irreps. Using the  $U(3\Omega) \supset U(\Omega)  \otimes
[U(3) \supset SO(3)]$ algebra, $m$ bosons states can be written as $\l|
m;\{f_1\, f_2\, f_3\}\, \alpha ; (\lambda\,\mu)\, K\; S\, M_S\ran$; The number
of $\alpha$ values is $d_{\Omega}(f_1\, f_2\, f_3)$, $K$ values follow from Eq.
(\ref{eq.bege-p15}) and $-S \leq M_S \leq S$. Note that $m$ and $(\lambda\,\mu)$
give a unique $\{f_1\, f_2\, f_3\}$. Therefore $H$-matrix dimension in
fixed-$(m,S)$ space is given by
\be
d(m,S) = \dis\sum_{\{f_1\, f_2\, f_3\} \in m} \;d_{\Omega}(f_1\, f_2\, f_3)\;
d_{(\lambda\,\mu)}(S)\;,
\label{eq.bege-p17}
\ee
and they will satisfy the sum rule $\sum_S\,(2S+1)d(m,S) ={3\Omega +m-1 \choose
m}$. Also, the dimension $D(m,M_S=0)$ of the $H$-matrix in the basis discussed
earlier is $D(m,M_S=0) = \sum_{S \in m} d(m,S)$.  For example, for  $(\Omega=4,
\;m=8)$, the dimensions $d(m,S)$ for $S=0-8$ are $714$, $1260$, $2100$, $1855$,
$1841$, $1144$, $840$, $315$ and $165$ respectively. Similarly, for
$(\Omega=6,\;m=10)$, the dimensions for $S=0-10$ are $51309$, $123585$,
$183771$, $189630$, $178290$, $133497$, $94347$, $51645$, $27027$, $9009$  and
$3003$ respectively. Because of these very large dimensions, numerical analysis
of BEGOE(1+2)-S1 ensemble is quite difficult.

\section{Results for spectral properties: propagation of energy centroids and
spectral variances}
\label{sec:s4}

\subsection{Eigenvalue density and NNSD: numerical results}

\label{sec:s4a}
\begin{figure}
\centering
\includegraphics[width=\linewidth]{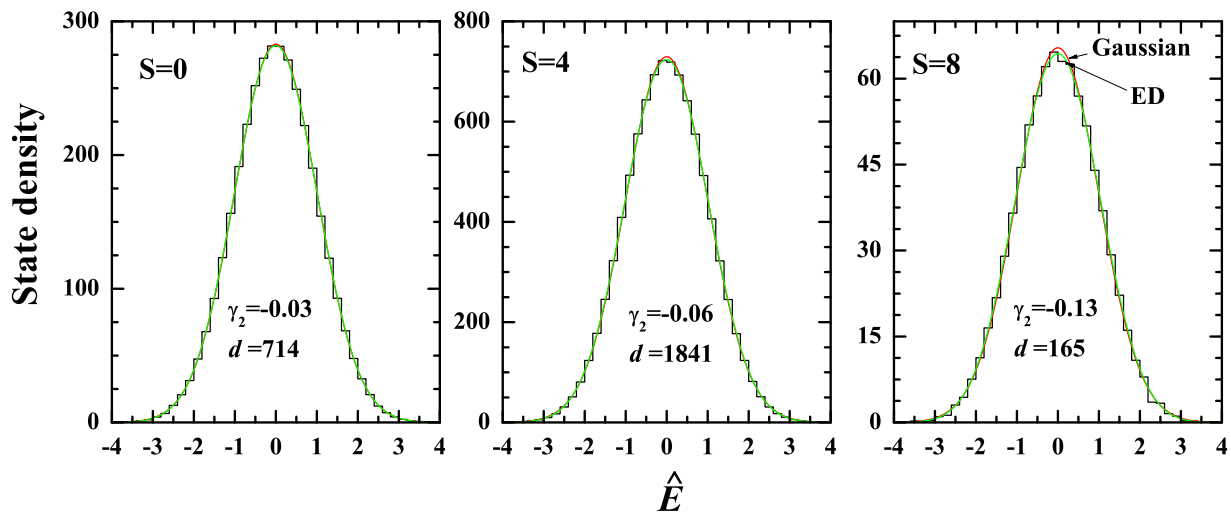}

\caption{Ensemble averaged eigenvalue density $\rho^{m,S}(\we)$ vs normalized
energy, $\we=\frac{E-E_c(m,S)}{\sigma}$, for a 100 member  BEGOE(2)-$S1$
ensemble with $\Omega=4$, $m=8$ and spin $S$=0, 4 and 8. The red curves give
Gaussian representation while the green curves are Edgeworth corrected Gaussians
(ED). The ensemble averaged values of excess $(\gamma_2)$ parameters are as
shown in figure. Note that Skewness $\gamma_1 \sim 0$ in all cases. In the
plots, the state densities, for a given spin $S$, are normalized to dimension
$d(m,S)$. Note that the total dimensionality of $H$-matrix here is $\sum_S d(m,S)=10234$.}

\label{den}
\end{figure}

Using the method described in Section \ref{sec:s2}, in some examples BEGOE(2)-$S1$ ensemble
has been constructed and analyzed are eigenvalue density and spectral
fluctuations.  Figure \ref{den} presents the results for the ensemble-averaged
fixed-($m$,$S$) eigenvalue density $\overline{\rho^{m,S}(E)}$ for the
BEGOE(2)-$S1$ ensemble defined by $h(1)=0$ in Eq. (\ref{eq.bege-p2}). We have
considered a 100-member BEGOE(2)-$S1$ ensemble with $m=8$ and $\Omega = 4$. The
strengths of the two-body interaction in the $s=0$, $s=1$ and $s=2$ channels are
chosen to be equal i.e. $\lambda_0=\lambda_1=\lambda_2$. In the construction of
the ensemble averaged eigenvalue densities, the spectra of each member of the
ensemble are first zero centered and scaled to unit width. The eigenvalues are
then denoted by $\we$. Given the fixed-($m$,$S$) eigenvalue centroids $E_c(m,S)$
and spectral widths $\sigma(m,S)$, $\we = [E-Ec(m,S)]/\sigma(m,S)$. Then the
histograms for the density are generated by combining the eigenvalues $\we$ from
all the members of the ensemble. In the figure, histograms are constructed with a
bin size equal to 0.2. Results are shown in Fig. \ref{den} for $S=0$, $4$ and
$8$ values. It is clearly seen that the eigenvalue densities are close to
Gaussian also the agreements with Edgeworth (ED) corrected Gaussians are excellent.

\begin{figure}
\centering
\includegraphics[width=\linewidth]{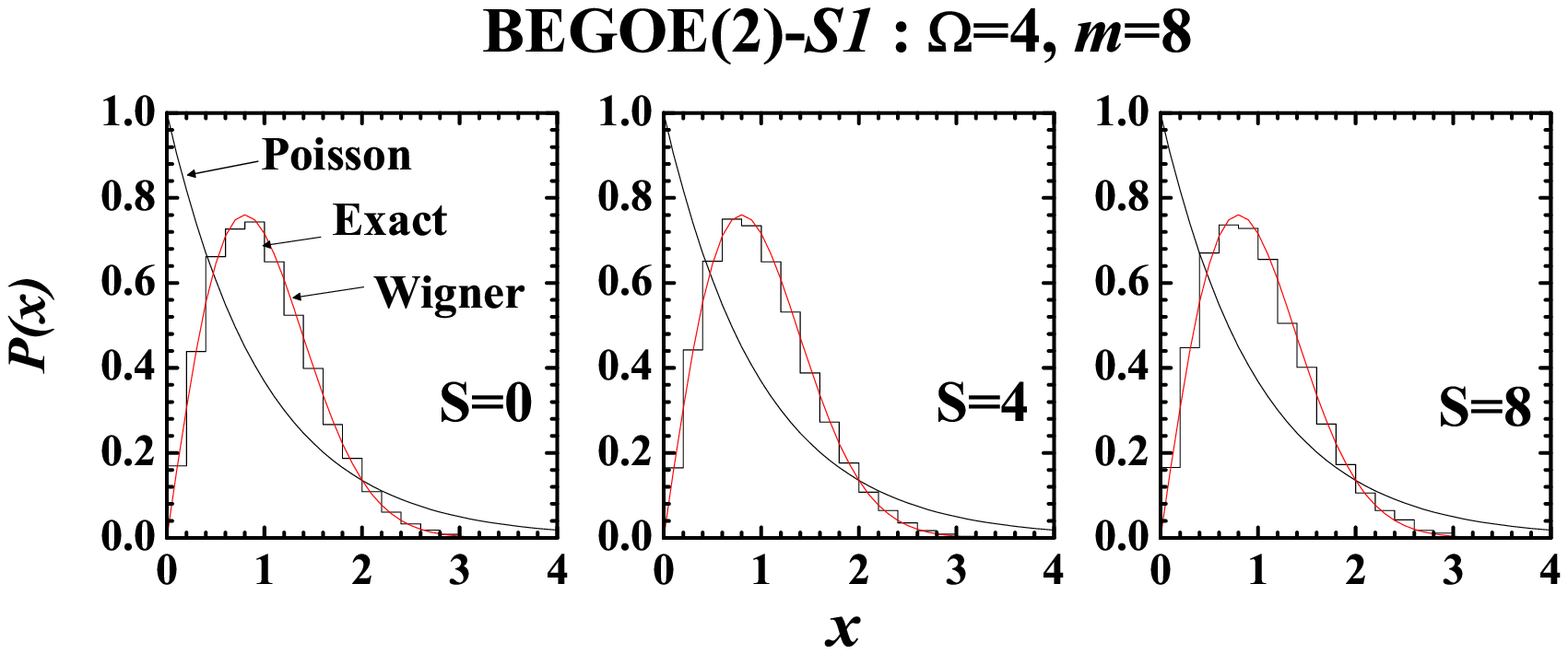}

\caption{Ensemble averaged Nearest Neighbor Spacing Distribution(NNSD)
histogram for a 100 member BEGOE(2)-$S1$ with $m=8$ and $\Omega=4$.   Results
are shown for the spin values $S$=0, 4 and 8. Here $x$ is in the units of local
mean spacing. Results are compared with Poisson and GOE (Wigner) forms.}

\label{nnsd}
\end{figure}

The nearest neighbor spacing distribution (NNSD), which gives information about
level repulsion, is of GOE type for spin-less BEGOE(2) \cite{Pa-00} and
BEGOE(2)-$F$ \cite{vyas-12}. In Fig. \ref{nnsd} NNSD results are shown for
BEGOE(2)-$S1$ with $m=8$ and $\Omega = 4$ for selected spin values. The NNSDs are
obtained by unfolding each spectrum in the ensemble, using the method described
in \cite{Pa-00}, with the smooth density as a corrected Gaussian with
corrections involving up to 6th order moments of the density function. In the
calculations, 80\% of the eigenvalues (dropping 10\% from both ends of the
spectrum) from each member are employed. It is clearly seen from the figures
that the NNSDs are close to the GOE (Wigner) form.

Previously it was shown that BEGOE(1+2) for spinless boson systems
\cite{Pa-00,Ch-03} and BEGOE(1+2)-$F$ for two species boson systems
\cite{vyas-12} generate Gaussian eigenvalue densities in the dense limit and
fluctuations follow GOE in absence of the mean-field. Therefore,
combining these with the results in Figs. \ref{den},\ref{nnsd},  we can conclude
that for finite  isolated interacting boson systems the eigenvalue density will
be generically of Gaussian form and fluctuations, in absence of the mean-field,
follow GOE. As discussed in \cite{Ch-03,vyas-12}, with mean-field, the
interaction strength has to be larger than a critical value for the fluctuations
to change from Poisson like to GOE.

\begin{table}[htp]
\caption{$\{f\}$, $(\lambda\,\mu)$ and $S$ labels for $m \leq 4$
bosons and the averages of $\hat{X}_3$ and $\hat{X}_4$ operators}
\begin{center}
\begin{tabular}{cccccc}
\hline
\\
$m$ & $\{f\}$ & $(\lambda\,\mu)$ & $S$ & $\lan X_3\ran$ & $\lan X_4\ran$ \\
\\
\hline
\\
$0$ & $\{0\}$ & $(00)$ & $0$ & $0$ & $0$ \\
$1$ & $\{1\}$ & $(10)$ & $1$ & $5$ & $-25$ \\
$2$ & $\{2\}$ & $(20)$ & $0$ & $0$ & $0$ \\
& & & $2$ & $21$ & $-147$ \\
& $\{11\}$ & $(01)$ & $1$ & $-5$ & $-25$ \\
$3$ & $\{3\}$ & $(30)$ & $1$ & $9$ & $-81$ \\
& & & $3$ & $54$ & $-486$ \\
& $\{21\}$ & $(11)$ & $1$ & $0$ & $-135$ \\
& & & $2$ & $0$ & $-81$ \\
& $\{111\}$ & $(00)$ & $0$ & $0$ & $0$ \\
$4$ & $\{4\}$ & $(40)$ & $0$ & $0$ & $0$ \\
& & & $2$ & $33$ & $-363$ \\
& & & $4$ & $110$ & $-1210$ \\
& $\{31\}$ & $(21)$ & $1$ & $-7$ & $-121$ \\
& & & $2$ & $21$ & $-459$ \\
& & & $3$ & $18$ & $-246$ \\
& $\{22\}$ & $(02)$ & $0$ & $0$ & $0$ \\
& & & $2$ & $-21$ & $-147$ \\
& $\{211\}$ & $(10)$ & $1$ & $5$ & $-25$ \\
\\
\hline	
\end{tabular}
\label{x3x4}
\end{center}
\end{table}

\subsection{Propagation of energy centroids and spectral variances}

\label{sec:s4b}
As the eigenvalue density is close to Gaussian, it is useful to derive
formulas for energy centroids and spectral variances in terms of sp energies
$\epsilon_i$ and the two-particle $V(2)$ matrix elements $V^s_{ijkl}$. They will
also allow us to study, numerically, fluctuations in energy centroids and
spectral variances. Simple propagation equation for the fixed-$(m,S)$ energy
centroids $\lan H\ran^{m,S}$ in terms of the scalars $\hat{n}$ and $S^2$
operators [their eigenvalues are $m$ and $S(S+1)$] is not possible. This is
easily seen from the fact that upto 2 bosons, we have 5 states ($m=0,S=0$;
$m=1,S=1$; $m=2,S=0,1,2$) but only 4 scalar operators ($1$, $\hat{n}$,
$\hat{n}^2$, $\hat{S}^2$). For the missing operator we can use $\hat{C}_2(SU(3)$
but then only fixed-$(m,(\lambda\,\mu) S)$ averages will propagate \cite{Ko-06a}.
The propagation equation is,
\be
\barr{l}
\lan \wh(1+2) \ran^{m, (\lambda\,\mu), S} = \lan \whh(1) + \wv(2)
\ran^{m, (\lambda,\mu), S} = m\;\lan \whh(1)\ran^{1,(10),1} \\
\\
+ \l[-\dis\frac{m}{6} + \dis\frac{m^2}{18} + \dis\frac{C_2(\lambda\,\mu)}{9} -
\dis\frac{S(S+1)}{6}\r]\; \;\lan \wv(2)\ran^{2,(20),0} \\
\\
+ \l[-\dis\frac{5m}{6} + \dis\frac{5m^2}{18} + \dis\frac{C_2(\lambda\,\mu)}{18}
+\dis\frac{S(S+1)}{6}\r]\; \;\lan \wv(2)\ran^{2,(20),2} \\
\\
+ \l[\dis\frac{m}{2} + \dis\frac{m^2}{6} - \dis\frac{C_2(\lambda\,\mu)}{6}
\r]\; \;\lan \wv(2)\ran^{2,(01),1}\;.
\earr \label{eq.bege-p18}
\ee
Now summing over all $(\lambda\,\mu)$ irreps that contain a given $S$ will give
$\lan \wh(1+2) \ran^{m,S}$. This is used to verify the codes we have developed
for constructing BEGOE(1+2)-$S1$ members. Propagation equation for spectral
variances $\lan [\wh(1+2)]^2 \ran^{m,S}$ is more complicated. Just as with
energy centroids, it is possible to propagate the variances $\lan [\wh(1+2)]^2
\ran^{m,(\lambda\,\mu),S}$. Towards this, first it should be noted that upto
$m=4$, there are $19$ states as shown in Table 1. Therefore, for propagation we
need $19$ $SO(3)$ scalars that are of maximum body rank $4$. For this the
invariants $\hat{n}$, $\hat{S}^2$, $\hat{C}_2(SU(3))$ and $\hat{C}_3(SU(3))$
will not suffice as they will give only 15 scalar operators. The missing three
operators can be constructed using the $SU(3) \supset SO(3)$ integrity basis
operators $\hat{X}_3$ and $\hat{X}_4$ that are $3-$ and $4-$body in nature; see
\cite{Ko-06a,DR-NPA}. Definition of these operators are given in \cite{DR-NPA}
and we call the operators given in this paper as $X^{DR}_3$ and $X^{DR}_4(k)$.
In the present work we have employed the following  definitions,
\be
\hat{X}_3 = -\dis\frac{5}{\dis\sqrt{10}}\,X^{DR}_3\;,\;\;\;\;
\hat{X}_4 = 5 \,X^{DR}_4(1)\;.
\label{eq.bege-p19}
\ee
Formulas for the averages $X_i((\lambda\,\mu),S) =  \lan \hat{X}_i \ran^{
(\lambda\,\mu),S}$ are given  by Eqs. (8)-(10) of \cite{DR-NPA} and they involve $SU(3) \supset
SO(3)$ reduced Wigner coefficients. Using the programmes for these, given in
\cite{DA-comp}, averages for $\hat{X}_3$ and $\hat{X}_4$ in the $19$ states with
$m \leq 4$ are calculated and the results are given in Table 1. Eqs.
(\ref{eq.bege-p13}) and (\ref{eq.bege-p13-1}) respectively will give
$C_2(\lambda\,\mu)$ and $C_3(\lambda\,\mu)$. Propagation equation for spectral
variances over fixed-$(\lambda\,\mu),S$ space can be written as,
\be
\barr{l}
\lan {\wh}^2 \ran^{m,(\lambda\,\mu),S} = \dis\sum_{i=1}^{19}\;a_i\;\cac_i\;;\\
\cac_1=1,\, \cac_2=m,\, \cac_3=m^2,\, \cac_4=m^3,\, \cac_5=m^4,\,
\cac_6=C_2(\lambda\,\mu)\,,\\
\cac_7=m\,C_2(\lambda\,\mu)\,,
\cac_8=m^2\,C_2(\lambda\,\mu),\,
\cac_9=S(S+1),\, \cac_{10}=m\,S(S+1)\,,\\
\cac_{11}=m^2\,S(S+1)\,,
\cac_{12}=S(S+1)\,C_2(\lambda\,\mu),\, \cac_{13}=[S(S+1)]^2\,,\\
\cac_{14}=
[C_2(\lambda\,\mu)]^2\,,\cac_{15}=C_3(\lambda\,\mu),\,
\cac_{16}=m\,C_3(\lambda\,\mu)\,,\\
\cac_{17}=X_3[(\lambda\,\mu),S],\,
\cac_{18}=m\,X_3[(\lambda\,\mu),S],\, \cac_{19}=X_4[(\lambda\,\mu),S]\;.
\earr \label{eq.bege-p20}
\ee
As we know $\lan \cac_i\ran^{m,(\lambda\,\mu),S}$ for $m \leq 4$, we can use
Using $\lan {\wh}^2 \ran^{m,(\lambda\,\mu),S}$ for $m \leq 4$ as inputs (they can be calculated by explicit construction of the Hamiltonian matrices using the method discussed in Section \ref{sec:s2})
one
can solve Eq. (\ref{eq.bege-p20}) to obtain the $a_i$'s. Then,  Eq.
(\ref{eq.bege-p20}) can be used to calculate $\lan {\wh}^2 \ran^{m,
(\lambda\,\mu),S}$ for any $m$, $(\lambda\,\mu)$ and $S$. However we still need to
evaluate numerically $X_3[(\lambda\,\mu),S]$ and $X_4[(\lambda\,\mu),S]$. Their
values are shown for $m=6$ and $8$ examples in Table 2. Spectral variances
$\lan \wh^2\ran^{m,S}$ over fixed-$S$ space can be obtained easily using $\lan
{\wh}^2 \ran^{m,(\lambda\,\mu),S}$. Let us add that there are methods
\cite{Wo-86},  though much more cumbersome, that will give directly $\lan
\wh^2\ran^{m,S}$. One such method is to use $(m_1, m_2, m_3)$
configurations introduced in Section \ref{sec:s2} and evaluate traces over these spaces.
Here, trace propagation is simple for both $H$ and $H^2$ averages and
also  $(m_1, m_2, m_3)$ configuration have a definite $M_S$ value.
Now, a subtraction procedure using $ \lan H^p\ran^{(m_1, m_2, m_3)}$,
$p=1,2$ will give fixed $(m,S)$ energy centroids and spectral variances. This
procedure is being implemented and results of this will be reported elsewhere.

\begin{figure}[thb]
\centering

\includegraphics[width=4in,height=3in]{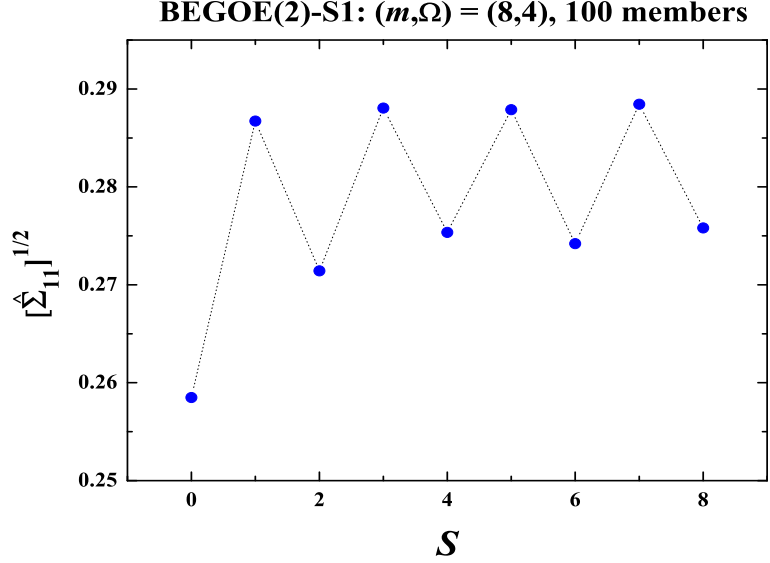}

\caption{$[\Sigma_{11}(m,S:m,S]^{1/2}$ giving width of the fluctuations in
energy centroids scaled to the spectrum width, as a function of spin $S$ for
BEGOE(2)-$S1$ with $(m=8,\Omega=4)$.}

\label{sig-11}
\end{figure}

\begin{figure}[thb]
\centering

\includegraphics[width=4in,height=3in]{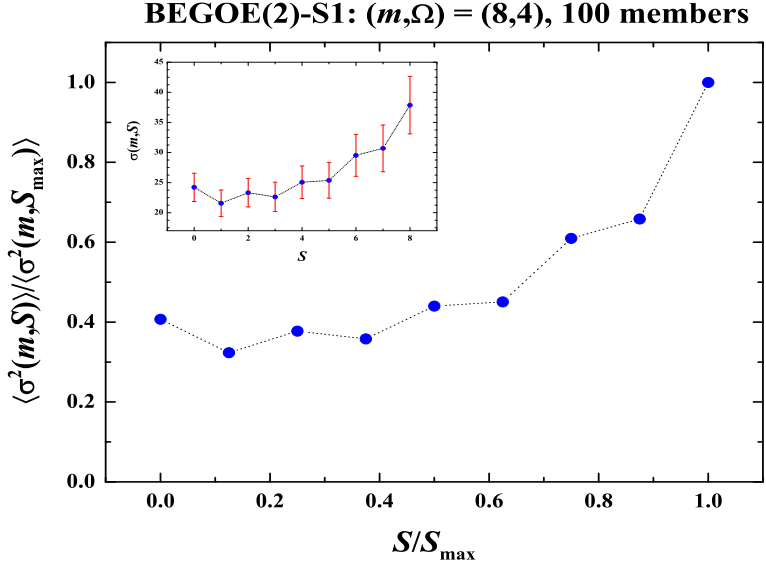}

\caption{Ensemble averaged variances scaled by that of maximum spin are shown
for BEGOE(2)-$S1$ ensembles with $(m=8,\Omega=4)$ as a function of $S/S_{max}$.
The inset figure represents ensemble averaged widths $\sigma(m,S)$ vs $S$ for
the same example. The r.m.s. deviations (over the ensemble) in the widths are
also shown as error bars.}

\label{width}
\end{figure}

Calculation of energy centroids and spectral variances for each member of the
ensemble will allow us to examine the covariances in these quantities. For
example,  normalized covariances in energy centroids is defined by
\be
\barr{rcl}
\Sigma_{11}(m,S:m^\pr,S^\pr) & = &
\dis\frac{\overline{\lan H \ran^{m,S}\lan H
\ran^{m^\pr,S^\pr}} - \l\{\overline{\lan H \ran^{m,S}}\r\} \;\;
\l\{\overline{\lan H \ran^{m^\pr,S^\pr}} \r\}
}{\dis\sqrt{\l\{\overline{\lan H^2 \ran^{m,S}}\r\}\;\;
\l\{\overline{\lan H^2 \ran^{m^\pr,S^\pr}}\r\}}}\;
\earr
\ee
For $(m,S) = (m^\pr,S^\pr)$ they will give information about fluctuations and in
particular about level motion in the ensemble \cite{Pa-00}. For $(m,S) \neq
(m^\pr,S^\pr)$, the covariances (cross correlations) are non-zero for BEGOE
while they will be zero for independent GOE representation for the $m$ boson
Hamiltonian matrices with different $m$ or $S$ (with fixed $\Omega$). We have
computed self-correlations $[\Sigma_{11}(m,S:m,S)]^{1/2}$ as a function of spin
$S$ for 100 member BEGOE(2)-$S1$ with $(m=8,\Omega=4)$ and the results are shown
in Fig. \ref{sig-11}. It is seen that the centroid fluctuations are large as
$[\Sigma_{11}]^{1/2} \sim 28\%$. However, the variation with spin $S$ is weak.
We have also calculated in some examples the variation of the average of
spectral variances with $S$ and the width of the fluctuations of the spectral
widths over the ensemble. Results are shown in Fig. \ref{width} for a
$(m=8,\Omega=4)$ system. It is clearly seen from the figure that the  variances
are almost constant for lower spins and increases for $S$ close to the maximum value
of $S$; a similar result is known for fermion systems \cite{KH}. Also, as seen
from the figure, the width of the fluctuations in spectral widths is much
smaller unlike the width of the fluctuations in energy centroids. Let us add
that near constancy of widths is a feature of many-body chaos
\cite{Ma-10epj,Fra-96}.

\begin{table}[htp]
\caption{$(\lambda\,\mu)$, $S$, $\hat{X}_3$ and $\hat{X}_4$ values for $m=6$
and $8$.}
\begin{center}
{\scriptsize{
\begin{tabular}{cccccccccccc}
\hline
\\
$m$ &$(\lambda\mu)$ & $S$ & $\hat{X}_3$ &  $\hat{X}_4$ &&& $m$ &$(\lambda\mu)$ & $S$ & $\hat{X}_3$ &  $\hat{X}_4$\\
\\
\hline
\\
				
$6$ & $(60)$ & $0$ & $0$ & $0$ 				&&&     $8$ & $(61)$ & $6$ & $375$ & $-7965$\\
    & $(60)$ & $2$ & $45$ & $-675$ 			&&&         & $(61)$ & $7$ & $340$ & $-5116$\\
    & $(60)$ & $4$ & $150$ & $-2250$ 			&&&         & $(42)$ & $0$ & $0$ & $0$ \\
    & $(60)$ & $6$ & $315$ & $-4725$ 			&&&         & $(42)$ & $2$ & $0$ & $-1131$ \\
    & $(41)$ & $1$ & $-9$ & $-297$ 			&&&         & $(42)$ & $3$ & $0$ & $-2046$ \\
    & $(41)$ & $2$ & $27$ & $-891$ 			&&&         & $(42)$ & $4$ & $91$ & $-2566$ \\
    & $(41)$ & $3$ & $36$ & $-702$ 			&&&         & $(42)$ & $5$ & $117$ & $-3567$ \\
    & $(41)$ & $4$ & $132$ & $-2466$ 			&&&         & $(42)$ & $6$ & $105$ & $-1701$ \\
    & $(41)$ & $5$ & $117$ & $-1431$ 			&&&         & $(50)$ & $1$ & $13$ & $-169$ \\
    & $(30)$ & $1$ & $9$ & $-81$ 			&&&         & $(50)$ & $3$ & $78$ & $-1014$ \\
    & $(30)$ & $3$ & $54$ & $-486$ 			&&&         & $(50)$ & $5$ & $195$ & $-2535$ \\
    & $(03)$ & $1$ & $-9$ & $-81$ 			&&&         & $(23)$ & $1$ & $-11$ & $-193$ \\
    & $(03)$ & $3$ & $-54$ & $-486$ 			&&&         & $(23)$ & $2$ & $33$ & $-1107$ \\
    & $(11)$ & $1$ & $0$ & $-135$ 			&&&         & $(23)$ & $3$ & $-33$ & $-1314$ \\
    & $(11)$ & $2$ & $0$ & $-81$ 			&&&         & $(23)$ & $4$ & $-44$ & $-1954$ \\
    & $(00)$ & $0$ & $0$ & $0$ 			&&&         & $(23)$ & $5$ & $-39$ & $-879$ \\
    & $(22)$ & $0$ & $0$ & $0$ 			&&&         & $(31)$ & $1$ & $-2$ & $-319$ \\
    & $(22)$ & $2$ & $0$ & $-603$ 			&&&         & $(31)$ & $2$ & $6$ & $-297$ \\
    & $(22)$ & $3$ & $0$ & $-990$ 			&&&         & $(31)$ & $3$ & $63$ & $-1164$ \\
    & $(22)$ & $4$ & $0$ & $-450$ 			&&&         & $(31)$ & $4$ & $55$ & $-640$ \\
	$8$ & $(80)$ & $0$ & $0$ & $0$            	&&&         & $(04)$ & $0$ & $0$ & $0$ \\
	    & $(80)$ & $2$ & $57$ & $-1083$ 	   	&&&         & $(04)$ & $2$ & $-33$ & $-363$ \\
	    & $(80)$ & $4$ & $190$ & $-3610$      	&&&         & $(04)$ & $4$ & $-110$ & $-1210$ \\
	    & $(80)$ & $6$ & $399$ & $-7581$      	&&&         & $(12)$ & $1$ & $7$ & $-121$ \\
	    & $(80)$ & $8$ & $684$ & $-12996$ 	 	&&&         & $(12)$ & $2$ & $-21$ & $-459$ \\
	    & $(61)$ & $1$ & $-11$ & $-553$       	&&&         & $(12)$ & $3$ & $-18$ & $-246$ \\
	    & $(61)$ & $2$ & $33$ & $-1467$       	&&&         & $(20)$ & $0$ & $0$ & $0$ \\
	    & $(61)$ & $3$ & $54$ & $-1398$       	&&&         & $(20)$ & $2$ & $21$ & $-147$ \\
	    & $(61)$ & $4$ & $166$ & $-3994$      	&&&         & $(01)$ & $1$ & $-5$ & $-25$ \\
	    & $(61)$ & $5$ & $171$ & $-2919$      	&&&         &&&&\\
\hline	
\end{tabular}
\label{m6m8}
}}
\end{center}
\end{table}				

\section{Pairing algebras and ground state structure}
\label{sec:s5}

In the  BEGOE(1+2)-$S1$ space, it is possible to identify two different pairing
algebras (each defining a particular type of pairing) and they follow from the
results in \cite{Ko-98,Ko-00,KoCa}. One of them corresponds to the $SO(\Omega)$
algebra in $U(3\Omega) \supset [U(\Omega) \supset SO(\Omega)] \otimes [U(3)
\supset SO(3)]$ and we refer to this as $SO(\Omega)-SU(3)$ pairing. The other
corresponds to the $SO(3\Omega)$ in $U(3\Omega) \supset SO(3\Omega)   \supset
SO(\Omega) \otimes SO(3)$. Note hat both the algebras have $SO(3)$ subalgebra
that generates the spin $S$. Here below we will give some details of these
pairing algebras. Inclusion of pairing Hamiltonians in BEGOE(1+2)-$S1$ $H$ will
alter the structure of ground states and this will be discussed in Section \ref{sec:s5c}.

\subsection{$SO(\Omega)-SU(3)$ pairing}
\label{sec:s5a}

Following the results given in \cite{Ko-98,Ko-00,KoCa} it is easy to identify
the $\Omega(\Omega-1)/2$ number of generators $U(i,j)$, $i<j$ of $SO(\Omega)$ in
$U(3\Omega) \supset [U(\Omega) \supset SO(\Omega)] \otimes [U(3)
\supset SO(3)]$,
\be
\barr{l}
U(i,j)= \dis\sqrt{\alpha(i,j)}\,\l[ g(i,j) + \alpha(i,j)\,g(j,i)\r]\,,
\;\;\;i<j\;;\\
\l|\alpha(i,j)\r|^2=1,\;\;\;\alpha(i,j)=\alpha(j,i),\;\;\;
\alpha(i,j)\alpha(j,k)=-\alpha(i,k)\;.
\earr \label{eq.bege-p21}
\ee
Note that $g(i,j)$ are defined in Eq. (\ref{eq.bege-p9}). The quadratic Casimir
invariant of $SO(\Omega)$ is,
\be
\hat{C}_2(SO(\Omega)) = \dis\sum_{i < j}\;U(i,j) \cdot U(j,i)\;.
\label{eq.bege-p22}
\ee
Applying Eq. (\ref{eq.bege-p21}) now gives,
\be
\barr{rcl}
\hat{C}_2(SO(\Omega)) & = & \dis\sum_{i<j}\,\alpha(i,j) \l[g(i,j) \cdot g(i,j) +
g(j,i) \cdot g(j,i) + 2 \alpha(i,j)\, g(i,j)g(j,i)\r] \\
& = & \dis\sum_{i \neq j} g(i,j) \cdot g(j,i) + \dis\sum_{i \neq j}
\alpha(i,j)\, g(i,j)  \cdot g(i,j) \\
& = & \hat{C}_2(U(\Omega)) - \dis\sum_{i,j} \beta_i \beta_j\, g(i,j)
\cdot g(i,j)\;;
\\
\beta_i \beta_j & = & -\alpha(i,j),\;\mbox{for}\;\;i \neq j,\;\;\;
\l|\beta_i\r|^2=1\;.\\
\earr \label{eq.bege-p23}
\ee
Here we have introduced $\beta_i$'s and the $\alpha(i,j)$ are defined in Eq.
(\ref{eq.bege-p21}). Now defining the pairing operator $\capp^k_q$, $k=0,2$
as
\be
\capp^k_q = \dis\sum_i\;\beta_i \l(b^\dg_{i;1} b^\dg_{i;1}\r)^k_q\;;\;\;\;
k=0,2 \\
\label{eq.bege-p24}
\ee
it is easy to see that,
\be
\barr{rcl}
H_\capp= \dis\sum_{k=0,2;q} \capp^k_q\,\l(\capp^k_q\r)^\dg & = &
\hat{C}_2(U(\Omega)) -
\hat{C}_2(SO(\Omega)) - \hat{n} \\
& = & \dis\frac{2}{3}\hat{C}_2(SU(3)) - \hat{C}_2(SO(\Omega)) -
(\Omega-4)\hat{n} + \dis\frac{\hat{n}^2}{3}\;.
\earr \label{eq.bege-p25}
\ee
In the final form above we have used Eqs. (\ref{eq.bege-p11-a}). Thus the
pairing Hamiltonian in the $U(3\Omega) \supset [U(\Omega) \supset SO(\Omega)]
\otimes [U(3) \supset SO(3)]$ algebra is a sum of $k=0$ and $2$ pairs and it is
simply related to the $SO(\Omega)$ and $SU(3)$ algebras. It is possible
enumerate the irreps of $SO(\Omega)$ given a $U(\Omega)$ or equivalently $SU(3)$
irrep (for a given $m$); see \cite{KoCa} and references therein. In terms of
these irrep labels and $SU(3)$ labels $(\lambda \, \mu)$, eigenvalues of
$H_\capp$ will follow from Eq. (\ref{eq.bege-p25}). This and the complimentary
non-compact $sp(6)$ pairing algebra generated by $\capp^k_q$,
$\l(\capp^k_q\r)^\dg$, $h^1_q$, $h^2_q$ and $\hat{n}$ will be discussed
elsewhere. For a recent review on complimentary Algebras see \cite{Rowe}.
Finally, the two-particle matrix elements of $H_\capp$ are $V_{iijj}^{s=0}=1$,
$V_{iijj}^{s=2}=1$ and all other matrix elements are zero.

Before going further, it is useful to mention that the Majorana operator
($\hat{M}$) that changes the space labels $(i,j)$ in a two-particle states
without changing the spin labels $m_\cs$ related in a simple manner to
$\hat{C}_2(U(3))$. Denoting the spin labels by $\alpha, \beta ,\ldots$, we have
\be
\hat{M}=\dis\sum_{i,j; \alpha,\beta}\;b^\dg_{j,\alpha}b^\dg_{i,\beta}
\l(b^\dg_{i,\alpha}b^\dg_{j,\beta}\r)^\dg = \hat{C}_2(U(3)) -3 \hat{n}\;.
\label{eq.bege-p25-a}
\ee

\subsection{$SO(3\Omega)$ pairing}

\label{sec:s5b}
Second pairing algebra follows from the recognition that $U(3\Omega)$ admits
$SO(3\Omega)$ subalgebra and as we will see ahead, the pairing here is generated
by $k=0$ pairs $b^\dg_{i} \cdot b^\dg_{i}$ alone. Following the results in
\cite{Ko-00} the generators of $SO(3\Omega)$ are easy to identify and they are,
\be
\barr{l}
u^{k=1}_q(i,i)\;;\;\;i=1,2,\ldots, \Omega\;, \\
V^k_q(i,j)= \dis\sqrt{(-1)^k\alpha(i,j)}\,\l[u^k_q(i,j) + \alpha(i,j)\,(-1)^k\;
u^k_q(j,i)\r]\,,
\;\;\;i<j\;;\\
\l|\alpha(i,j)\r|^2=1,\;\;\;\alpha(i,j)=\alpha(j,i),\;\;\;
\alpha(i,j)\alpha(j,k)=-\alpha(i,k)\;.
\earr \label{eq.bege-p26}
\ee
The operators $u^k_q$ are defined by Eq. (\ref{eq.bege-p4}). Carrying out
angular momentum algebra the following relation between the quadratic Casimir
invariants $\hat{C}_2(SO(3\Omega))$ and $\hat{C}_2(U(3\Omega))$,
of $SO(\Omega)$ and $U(3\Omega))$, can be established using
Eqs. (\ref{eq.bege-p26}) and (\ref{eq.bege-p5}),
\be
\barr{l}
\hat{C}_2(SO(3\Omega)) = 2\dis\sum_i u^1(i,i) \cdot u^1(i,i) +\dis\sum_{i<j;k}
V^k(i,j) \cdot V^k(i,j) \\
= \hat{C}_2(U(3\Omega)) - \dis\sum_{i,k} (-1)^k u^k(i,i) \cdot
u^k(i,i) + \dis\sum_{i \neq j; k} (-1)^k \alpha(i,j) u^k(i,j) \cdot
u^k(i,j)\;.
\earr \label{eq.bege-p27}
\ee
Introducing the pairing operator $P_+$,
\be
P_+ = \dis\sum_{i} \gamma_i\, P_+(i) = \dis\frac{1}{2} \dis\sum_{i} \gamma_i\,
b^\dg_{i;1} \cdot b^\dagger_{i;1} \;,\;\;\;P_-=(P_+)^\dg
\label{eq.bege-p28}
\ee
we can prove the following relationship between $\hat{C}_2(SO(3\Omega))$ and the
pairing Hamiltonian $H_P=4P_+P_-$,
\be
\barr{rcl}
4\,H_P = 4P_+P_- & = & -\hat{n} + \hat{C}_2(U(3\Omega)) - \hat{C}_2(SO(3\Omega))
\\
& = & \hat{n}(\hat{n} + 3\Omega-2) -\hat{C}_2(SO(3\Omega)) \;;\\
\gamma_i \gamma_j & = & -\alpha(i,j),\;\mbox{for}\;\;i \neq j,\;\;\;
\l|\gamma_i\r|^2=1\;.
\earr \label{eq.bege-p29}
\ee
The $\beta \leftrightarrow \alpha$ relation is needed for the correspondence
between $H_P$ and $\hat{C}_2(SO(3\Omega))$. Important point now being that the
three operators $P_+$, $P_-$ and $P_0=(\Omega + \hat{n})/2$ will form a
$SU(1,1)$ algebra complimentary to $SO(3\Omega)$. Thus the $SO(3\Omega)$ pairing
is much simpler. With $U(3\Omega))$ irreps being $\{m\}$, the $SO(3\Omega)$
irreps are labeled by  the seniority quantum number $\omega$ where,
\be
\omega = m, m-2, \ldots,0\;\;\mbox{or}\;\;\;1\;.
\label{eq.bege-p30}
\ee
and $H_P$ eigenvalues are
\be
\lan H_P\ran^{m,\omega} = \dis\frac{1}{4} (m-\omega)(m+\omega+3\Omega-2)\;.
\label{eq.bege-p31}
\ee
The two particle matrix elements of $H_P$ are simply $V^{s=0}_{iijj}=1$ and all
other matrix elements are zero.

\subsection{Ground state structure}

\label{sec:s5c}
\begin{figure}
\centering

\includegraphics[width=5in,height=3in]{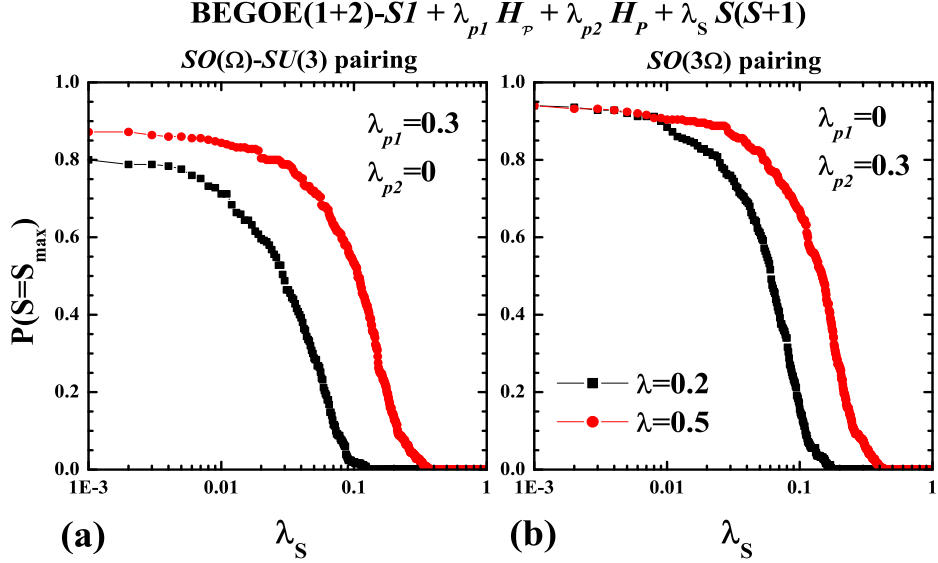}

\caption{(a) Probability for ground state to have maximum spin as a function of
$\lambda_S$ for a 250 member $(m=6,\Omega=4)$  BEGOE(1+2)-$S1$ system with $H$
defined by Eq. (\ref{eq.bege-p32}). In all the calculations
$\lambda_0=\lambda_1=\lambda_2=\lambda$ and the results are shown for
$\lambda=0.2$ and $0.5$. (a) results with $SO(\Omega)-SU(3)$ pairing parameter
$\lambda_{p1}=0.3$ and $SO(3\Omega)$ pairing parameter $\lambda_{p2}=0$ in Eq.
(\ref{eq.bege-p32}). (b) Same as (a) but with $SO(\Omega)-SU(3)$ pairing
parameter $\lambda_{p1}=0$ and $SO(3\Omega)$ pairing parameter $\lambda_{p2}
=0.3$ in Eq. (\ref{eq.bege-p32}).}

\label{p-smax}
\end{figure}

\begin{figure}
\centering

\includegraphics[width=4in,height=5.5in]{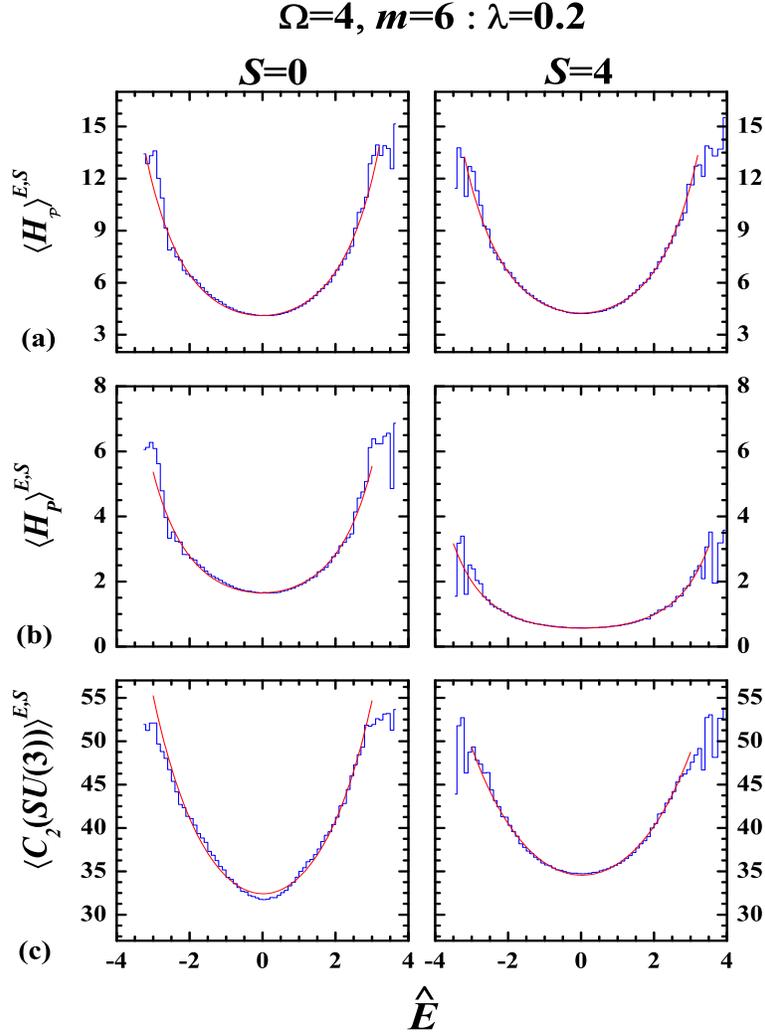}

\caption{Expectation values of the two pairing Hamiltonians and $C_2(SU(3))$
 vs $\we$ for a 250 member BEGOE(1+2)
systems with $H$ defined by Eq.  (\ref{eq.bege-p2}) and $(\Omega=4,m=6)$.
Results are shown for spins  $S=0$ and $S=4$. (a) expectation values of
$H_\capp$, (b) expectation values of $H_P$ and (c) expectation value of
$C_2(SU(3))$. Ensemble averaged results are shown by histograms while continuous curves are ratio
of Gaussians given by EGOE theory \cite{vyas-12}. See text for further details.}

\label{expect}
\end{figure}

With two different pairings in the BEGOE(1+2)-$S1$ space, analysis of properties
of spin one boson systems with the following extended Hamiltonian $H_{ext}$
will be interesting and useful,
\be
\barr{rcl}
\l\{\wh_{ext}\r\} & = & \whh(1) + \lambda_0
\l\{\wv^{s=0}(2)\r\} + \lambda_1 \l\{\wv^{s=1}(2)\r\} +
\lambda_2 \l\{\wv^{s=2}(2)\r\} \\
& & + \lambda_{p1} H_\capp + \lambda_{p2} H_P + \lambda_S\;{\hat S}^2\;.
\earr \label{eq.bege-p32}
\ee
As an example, shown in Fig. \ref{p-smax} are the results for the probability
that the ground state spin is $S=S_{max}=m$. It is seen that even with strong
random interaction the probability is not $100\%$ and with increasing
$\lambda_S$, the probability rapidly comes down to zero. The situation here is
different from the result seen for BEGOE(1+2)-$F$ where strong enough
interaction generates states with maximum spin with 100\% probability. With
$SO(3\Omega)$ pairing the drop in probability is more rapid compared to the
situation with $SO(\Omega)-SU(3)$ pairing. In addition, we have also calculated
the expectation values of the two pairing Hamiltonians and $C_2(SU(3))$ and the
results are shown in Fig. \ref{expect}. We have considered BEGOE(1+2)
Hamiltonian defined by Eq. (\ref{eq.bege-p2}) with $\lambda_0 = \lambda_1 =
\lambda_2 = \lambda = 0.2$, i.e. in the region of chaos generated by random
two-body interactions in the presence of a mean-field. It is seen that the
expectation values are largest near the ground states and then decrease as we
move towards the center of the spectrum. Due to finiteness of the model space,
the curves are essentially symmetric around the center. The calculated results
are in good agreement with the prediction \cite{vyas-12} that for boson systems
(just as it was well verified for fermion systems), expectation values will be
ratio of Gaussians; see Section \ref{sec:s6} and Eq. (43) in \cite{vyas-12}. Results in the
figure show that with repulsive pairing, ground states will be dominated by low
seniority structure. Also, with random interactions, there is no clear
distinction between the two different pairing structures. Variation with the
$\lambda$ parameter for larger systems ($\Omega$ and $m$ large)  may show the
difference but these calculations  are not attempted as  the matrix dimensions
will be very large. This exercise will be attempted in future.


\section{Conclusions and Future outlook}

\label{sec:s6}
Introduced in this paper is the embedded Gaussian orthogonal ensemble of  random
matrices generated by random two-body interactions in presence of a mean-field
for spin one boson systems. Presented are some first analytical and numerical
results for this ensemble. Due to large fixed-$(m,M_S=0)$ matrix dimensions,
only restricted numerical calculations (with dimensions less than 10000) could
be carried out at present. Some results for spectral properties including the
form of eigenvalue density close to Gaussian, NNSD following GOE for
sufficiently strong interaction strength and also for lowest two moments of the
two point function are presented. It is possible to deal with much larger space
if we use direct construction of $H$ matrix in a good $S$ basis. This is being
attempted and using this in future a more detailed study with much larger size
examples will be reported. Preliminary aspects of one of  the embedding algebras
$SU(\Omega) \otimes SU(3)$ and also two pairing algebras in the space defining
BEGOE(1+2)-$S1$  are discussed in the paper. More detailed study of  the effects
of random interactions in presence of the two pairing interactions will be
discussed elsewhere. Extension of BEGOE(2)-$S1$ to BEGUE(2)-$S1$ and to the more
restricted BEGUE(2)-$SU(3)$ with $H$ preserving $SU(3)$ symmetry for spin one
boson systems are possible; see \cite{MK-12} for preliminary results for
BEGUE(2)-$SU(3)$. Finally, applications of BEGOE(1+2)-$S1$ ensemble to spin one
BEC should be possible in future.

\section*{Acknowledgments} Authors (N.D.C. and V.P.) acknowledge support from
UGC(New Delhi) grant F.No:40- 425/2011(SR) and No.F.6-17/10(SA-II) respectively.
M. V.  gratefully acknowledges financial
support from the US National Science Foundation grant PHY-0855337.

\ed